\documentclass[epj]{svjour}
\usepackage{amsmath}

\usepackage{graphicx}
\usepackage{fancyhdr}

\def\mytitle{Low energy particle physics and NLSUSY GR} 
\def\myauthors{K. Shima, M. Tsuda and W. Lang}  
\def\mytype{Cosmology and Astrophysics}
\def\mysession{}

\begin{document}
\title{Low energy particle physics 
and cosmology of nonlinear supersymmetric general relativity}
\author{
K. Shima\inst{1}
\thanks{\emph{Email:} shima@sit.ac.jp}%
 \and
M. Tsuda\inst{1}
\thanks{\emph{Email:} tsuda@sit.ac.jp}%
 \and
W. Lang\inst{2}
\thanks{\emph{Email:} wolfdieter.lang@physik.uni-karlsruhe.de}%
}
%
%
\institute{
Laboratory of Physics, Saitama Institute of Technology, 
Fukaya, Saitama 369-0293, Japan
 \and 
Institute for Theoretical Physics, 
Karlsruhe University, D-76128 Karlsruhe, Germany
}
%
\date{}
\abstract{
We show a low energy physical meaning of nonlinear supersymmetric general relativity 
(NLSUSY GR) in asymptotic Riemann-flat space-time 
by studying the vacuum structure of $N = 2$ linear supersymmetry (LSUSY) invariant QED, 
which is equivalent to $N = 2$ NLSUSY model, in two dimensional space-time. 
Two different vacuum field configurations of $SO(3,1)$ isometry describe the two different physical vacua, 
i.e. one breaks spontaneously both $U(1)$ and SUSY and the other breaks spontaneously SUSY alone, 
where the latter elucidates the mysterious relations between the cosmology and the (low energy) 
particle physics and gives a new insight into the origin of mass. 
} 
\maketitle
%

\noindent
By extending the geometric arguments of Einstein general relativity (EGR) 
on Riemann space-time to new space-time just inspired by nonlinear supersymmetry (NLSUSY), 
where tangent space-time is specified by not only $x_a$ for $SO(3,1)$ 
but also the Grassmanian $\psi_\alpha$ for isomorphic $SL(2,C)$ of NLSUSY, 
the fundamental action (called nonlinear supersymmetric general relativity (NLSUSY GR)) 
has been constructed \cite{KS2,KS3}, 
%
\begin{eqnarray}
& & L(w) = {c^4 \over 16{\pi}G} \vert w \vert (\Omega(w) - \Lambda), 
\label{SGM}
\\[2mm]
& & \hspace*{7mm} 
\vert w \vert = \det w^a{}_\mu = \det(e^a{}_\mu + t^a{}_\mu(\psi)), 
\nonumber \\[.5mm]
& & \hspace*{7mm} 
t^a{}_\mu(\psi) = {\kappa^2 \over 2i} 
(\bar\psi^i \gamma^a \partial_\mu \psi^i 
- \partial_\mu \bar\psi^i \gamma^a \psi^i), 
\end{eqnarray}
where $G$ is the Newton gravitational constant, 
$\Lambda$ is a ({\it small}) cosmological term and $i = 1, 2, \cdots, N$. 
$w^a{}_\mu(x)$ $= e^a{}_\mu + t^a{}_\mu(\psi)$, $e^a{}_\mu$ for the local $SO(3,1)$, 
$t^a{}_\mu(\psi)$ for the local $SL(2,C)$ and $\Omega(w)$ 
are the invertible unified vierbein of new spacetime, the ordinary vierbein of EGR, 
the stress-energy-momentum of superons $\psi(x)$ 
and the the unified scalar curvature of new (SGM) spacetime, respectively. 
$s_{\mu \nu} \equiv w^a{}_\mu \eta_{ab} w^b{}_\nu$ and 
$s^{\mu \nu}(x) \equiv w^\mu{}_a(x) w^{\nu a}(x)$ 
are unified metric tensors of SGM spacetime. 
New space-time is the generalization of the compact isomorphic groups $SU(2)$ and $SO(3)$ 
for the gauge symmetry of 't Hooft-Polyakov monopole 
into the noncompact isomorphic groups $SO(1,3)$ and $SL(2C)$ for space-time symmetry. 
NLSUSY GR action (\ref{SGM}) possesses promissing large symmetries 
isomorphic to $SO(N)$ ($SO(10)$) super-Poincar\'e group \cite{ST3,ST4}; 
namely, $L(w)$ is invariant under 
\begin{eqnarray}
& & [{\rm new \ NLSUSY}] \otimes [{\rm local \ GL(4,R)}] 
\nonumber \\[.5mm]
& & \otimes [{\rm local \ Lorentz}] \otimes \ [{\rm local \ spinor \ translation}] 
\end{eqnarray}
for spacetime symmetries 
and 
\begin{equation}
[{\rm global} SO(N)] \otimes [{\rm local} U(1)^N] 
\end{equation}
for the internal symmetries. 
Note that the no-go theorem is overcome (circumvented) in the sense that 
the non-tivial $N$-extended SUSY gravity theory with $N > 8$ has been constructed in the NLSUSY invariant way. 
NLSUSY GR $L(w)$ (called superon-graviton model (SGM) from composite viewpoint) on new {\it emp- ty} space-time 
written in the form of the {\it vacuum} Einstein - Hilbert (EH) type 
is unstable due to NLSUSY structure of tangent space-time and 
decays (called {\it Big Decay} \cite{ST4}) spontaneously into ordinary EH action 
with the cosmplogical constant $\Lambda$, NLSUSY action for $N$ Nambu-Goldstone (NG) fermions 
(called {\it superons} as hypothetical spin $1/2$ objects constituting all observed particles) 
and their gravitational interactions on ordinary Riemann space-time written formally 
as the following SGM action, which ignites Big Bang of the present observed universe; 
\begin{equation}
L(e,\psi)={c^{4} \over 16{\pi}G}\vert e \vert \{ R(e) - \Lambda + \tilde T(e, \psi) \},
\label{SGMR}
\end{equation}
%
%
where $R(e)$ is the scalar curvature of EH action and 
$\tilde T(e,\psi)$ represents the kinetic term and the gravitational interaction of superons. 
%

%
Considering that SGM action reduces to $N$ - extended NLSUSY action 
with ${\kappa^{2} = ({c^{4}\Lambda \over 8{\pi}G}})^{-1}$ in asymptotic 
Riemann-flat $(e^a{}_\mu \rightarrow \delta^a_\mu)$ space-time after Big Decay, 
it is interesting from the low energy physics viewpoints to construct the $N$-extended 
linear (L) SU- SY theory equivalent to $N$-extended NLSUSY model. 
We have shown explicitly by the heuristic arguments for simplicity 
in two space-time dimensions ($d = 2$) \cite{ST2,ST5} 
that $N = 2$ LSUSY interacting QED is equivalent (in a sense that SUSY invariant relations 
between basic fields in LSUSY theories and the NG fermions hold) to $N = 2$ NLSUSY model. 
(Note that the minimal realistic SUSY QED in SGM composite scenario is described by $N = 2$ SUSY \cite{STT2}.) 
%

Indeed, $N = 2$ NLSUSY action for two superons (NG fermions) $\psi^i\ (i=1,2)$ in $d = 2$ is written 
as follows, 
\begin{eqnarray}
& & L_{N=2{\rm NLSUSY}} 
\nonumber \\[.5mm]
& & = -{1 \over {2 \kappa^2}} \vert w \vert
\nonumber \\[.5mm]
& & = - {1 \over {2 \kappa^2}} 
\left\{ 1 + t^a{}_a + {1 \over 2!}(t^a{}_a t^b{}_b - t^a{}_b t^b{}_a) 
\right\} 
\nonumber \\[.5mm]
& & = - {1 \over {2 \kappa^2}} 
\left\{ 1 - i \kappa^2 \bar\psi^i \!\!\not\!\partial \psi^i \right. 
\nonumber \\[.5mm]
& & 
\left. - {1 \over 2} \kappa^4 
( \bar\psi^i \!\!\not\!\partial \psi^i \bar\psi^j \!\!\not\!\partial \psi^j 
- \bar\psi^i \gamma^a \partial_b \psi^i \bar\psi^j \gamma^b \partial_a \psi^j ) 
\right\} 
\label{VAaction2}
\end{eqnarray}
where $\kappa$ is a constant whose dimension is $({\rm mass})^{-1}$ and 
%
$\vert w \vert = \det(w^a{}_b) = \det(\delta^a_b + t^a{}_b)$, 
$t^a{}_b = - i \kappa^2 \bar\psi^i \gamma^a \partial_b \psi^i$. 
%
The most general $N = 2$ LSUSY QED action in $d = 2$, 
is written as follows for the massless case, 
\begin{eqnarray}
& & L_{N=2{\rm SUSYQED}} 
\nonumber \\[.5mm]
& & = - {1 \over 4} (F_{ab})^2 
+ {i \over 2} \bar\lambda^i \!\!\not\!\partial \lambda^i 
+ {1 \over 2} (\partial_a A)^2 
+ {1 \over 2} (\partial_a \phi)^2 
+ {1 \over 2} D^2 
\nonumber \\[.5mm]
& & 
- {1 \over \kappa} \xi D 
+ {i \over 2} \bar\chi \!\!\not\!\partial \chi 
+ {1 \over 2} (\partial_a B^i)^2 
+ {i \over 2} \bar\nu \!\!\not\!\partial \nu 
+ {1 \over 2} (F^i)^2 
\nonumber \\[.5mm]
& & 
+ f ( A \bar\lambda^i \lambda^i + \epsilon^{ij} \phi \bar\lambda^i \gamma_5 \lambda^j 
+ A^2 D - \phi^2 D - \epsilon^{ab} A \phi F_{ab} ) 
\nonumber \\[.5mm]
& & 
+ e \left\{ i v_a \bar\chi \gamma^a \nu 
- \epsilon^{ij} v^a B^i \partial_a B^j 
+ \bar\lambda^i \chi B^i 
+ \epsilon^{ij} \bar\lambda^i \nu B^j 
\right. 
\nonumber \\[.5mm]
& & 
\left. 
- {1 \over 2} D (B^i)^2 
+ {1 \over 2} (\bar\chi \chi + \bar\nu \nu) A 
- \bar\chi \gamma_5 \nu \phi \right\}
\nonumber \\[.5mm]
& & 
+ {1 \over 2} e^2 (v_a{}^2 - A^2 - \phi^2) (B^i)^2. 
\label{L2action}
\end{eqnarray}
where  $(v^a, \lambda^i, A, \phi, D)$ ($F_{ab} = \partial_a v_b - \partial_b v_a$) 
is the off-shell vector supermultiplet containing $v^a$ for a $U(1)$ vector field, 
$\lambda^i$ for doublet (Majorana) fermions 
and $A$ for a scalar field in addition to $\phi$ for another scalar field 
and $D$ for an auxiliary scalar field, 
while ($\chi$, $B^i$, $\nu$, $F^i$) is off-shell scalar supermultiplet containing 
$(\chi, \nu)$ for two (Majorana) are fermions, 
$B^i$ for doublet scalar fields and $F^i$ for auxiliary scalar fields. 
Also $\xi$ is an arbitrary demensionless parameter giving a magnitude of SUSY breaking mass, 
and $f$ and $e$ are Yukawa and gauge coupling constants with the dimension (mass)$^1$, respectively. 
$N = 2$ LSUSY QED action (\ref{L2action}) can be rewritten as the familiar manifestly covariant form 
wh- ich is manifestly invariant under the local $U(1)$ transformation. 
(For further details see ref.\cite{ST5}.)  

For extracting the low energy particle physics contents of $N = 2$ SGM (NLSUSY GR) 
we consider in Riemann-flat asymptotic space-time, where $N = 2$ SGM reduces to 
essentially $N = 2$ NLSUSY action equivalent to $N = 2$ SUSY QED action, i.e. 
%
\begin{eqnarray}
& & L_{N=2{\rm SGM}} \overset{e^a{}_\mu \rightarrow \delta^a_\mu}{\longrightarrow} 
\nonumber \\[.5mm]
& & 
L_{N=2{\rm NLSUSY}} + [{\rm suface\ terms}]
= L_{N=2{\rm SUSYQED}}. 
\end{eqnarray}
%
The equivalence of the two theories are shown explicitly by substituting 
the following generalized SUSY invariant relations \cite{ST5} into the LSUSY theory. 
The SUSY invariant relations for $(v^a, \lambda^i, A, \phi, D)$ 
as composites of $\psi^i$ are 
\begin{eqnarray}
& & 
v^a = - {i \over 2} \xi \kappa \epsilon^{ij} 
\bar\psi^i \gamma^a \psi^j \vert w \vert, 
\nonumber \\[.5mm]
& & 
\lambda^i = \xi \left[ \psi^i \vert w \vert 
- {i \over 2} \kappa^2 \partial_a 
\{ \gamma^a \psi^i \bar\psi^j \psi^j 
(1 - i \kappa^2 \bar\psi^k \!\!\not\!\partial \psi^k) \} \right], 
\nonumber \\[.5mm]
& & 
A = {1 \over 2} \xi \kappa \bar\psi^i \psi^i \vert w \vert, 
\nonumber \\[.5mm]
& & 
\phi = - {1 \over 2} \xi \kappa \epsilon^{ij} \bar\psi^i \gamma_5 \psi^j 
\vert w \vert, 
\nonumber \\[.5mm]
& & 
D = {\xi \over \kappa} \vert w \vert 
- {1 \over 8} \xi \kappa^3 
\partial_a \partial^a ( \bar\psi^i \psi^i \bar\psi^j \psi^j ), 
\end{eqnarray}
while for $(\chi, B^i, \nu, F^i)$, 
\begin{eqnarray}
& & 
\chi = \xi^i \left[ \psi^i \vert w \vert 
+ {i \over 2} \kappa^2 \partial_a 
\{ \gamma^a \psi^i \bar\psi^j \psi^j 
(1 - i \kappa^2 \bar\psi^k \!\!\not\!\partial \psi^k) \} \right], 
\nonumber \\[.5mm]
& & 
B^i = - \kappa \left( {1 \over 2} \xi^i \bar\psi^j \psi^j 
- \xi^j \bar\psi^i \psi^j \right) \vert w \vert, 
\nonumber \\[.5mm]
& & 
\nu = \xi^i \epsilon^{ij} \left[ \psi^j \vert w \vert 
+ {i \over 2} \kappa^2 \partial_a 
\{ \gamma^a \psi^j \bar\psi^k \psi^k 
(1 - i \kappa^2 \bar\psi^l \!\!\not\!\partial \psi^l) \} \right], 
\nonumber \\[.5mm]
& & 
F^i = {1 \over \kappa} \xi^i \left\{ \vert w \vert 
+ {1 \over 8} \kappa^3 
\partial_a \partial^a ( \bar\psi^j \psi^j \bar\psi^k \psi^k ) 
\right\} 
\nonumber \\[.5mm]
& & 
- i \kappa \xi^j \partial_a ( \bar\psi^i \gamma^a \psi^j \vert w \vert ) 
- {1 \over 4} e \kappa^2 \xi \xi^i \bar\psi^j \psi^j \bar\psi^k \psi^k, 
\label{SSUSYinv}
\end{eqnarray}
where $\xi^i$ in are arbitrary parameters satisfying $\xi^2 - (\xi^i)^2 = 1$. 
The familiar LSUSY transformations on the component fields of the supermultiplet are 
reproduced in terms of the NLSUSY transformations on the superons $\psi^i$ contained. \\
It is interesting that the four-fermion self-interaction term 
(i.e. the condensation of $\psi^i$) appearing in only the auxiliary fields $F^i$ 
is the origin of the familiar local $U(1)$ gauge symmetry of LSUSY theory. 
Is the condensation of superons the origin of the local gauge interaction? \\
We can show that the above arguments in the relation between $N = 2$ NLSUSY model 
and $N = 2$ LSUSY QED also hold for the theory 
constructed from {\it extended} scalar (matter) supermultiplets 
coupled to the vector supermultiplet. 

Now we study the vacuum structure of $N = 2$ SUSY QED action (\ref{L2action}) 
\cite{STL}. 
The vacuum is determined by the minimum of the potential $V(A, \phi, B^i, D)$, 
\begin{eqnarray}
& & V(A, \phi, B^i, D) 
\nonumber \\[.5mm]
& & =  - {1 \over 2} D^2 + \left\{ {\xi \over \kappa} 
- f(A^2 - \phi^2) + {1 \over 2} e (B^i)^2 \right\} D.  
\end{eqnarray}
Substituting the solution of the equation of motion for the auxiliary field $D$
we obtain 
%
\begin{equation}
V(A, \phi, B^i) = {1 \over 2} f^2 \left\{ A^2 - \phi^2 - {e \over 2f} (B^i)^2 
- {\xi \over {f \kappa}} \right\}^2 \ge 0. 
\end{equation}
%
The configurations of the fields corresponding to the vacua in $(A, \phi, B^i)$-space, 
which are $SO(1,3)$ or $SO(3,$ $1)$ invariant,   
are classified according to the signatures of the parameters $e, f, \xi, \kappa$ 
as follows: \\
(I) For $ef > 0$, \ \ ${\xi \over {f \kappa}} > 0$ case, 
%
\begin{equation}
A^2 - \phi^2 - (\tilde B^i)^2 = k^2. 
\ \left( \tilde B^i = \sqrt{e \over 2f} B^i, 
\ k^2 = {\xi \over {f \kappa}} \right) 
\end{equation}
%
(II) For $ef < 0$, \ \ ${\xi \over {f \kappa}} > 0$ case, 
%
\begin{equation}
A^2 - \phi^2 + (\tilde B^i)^2 = k^2. 
\ \left( \tilde B^i = \sqrt{-{e \over 2f}} B^i, 
\ k^2 = {\xi \over {f \kappa}} \right) 
\end{equation}
%
(III) For $ef > 0$, \ \ ${\xi \over {f \kappa}} < 0$ case, 
%
\begin{equation}
- A^2 + \phi^2 + (\tilde B^i)^2 = k^2. 
\ \left( \tilde B^i = \sqrt{e \over 2f} B^i, 
\ k^2 = - {\xi \over {f \kappa}} \right) 
\end{equation}
%
(IV) For $ef < 0$, \ \ ${\xi \over {f \kappa}}< 0$ case, 
%
\begin{equation}
- A^2 + \phi^2 - (\tilde B^i)^2 = k^2. 
\ \left( \tilde B^i = \sqrt{-{e \over 2f}} B^i, 
\ k^2 = - {\xi \over {f \kappa}} \right) 
\end{equation}
%
We find that the vacua (I) and (IV) with $SO(1,3)$ isometry in $(A, \phi, B^i)$-space are unphysical, 
for they produce the pathological wrong sign kinetic terms for the fields induced around the vacuum. 

As for the cases (II) and (III) we perform the similar arguments as shown below 
and find that two different physical vacua appear. 
%
%
%
The physical particle spectrum is obtained by expanding the field $(A, \phi, B^i)$ around the vacuum 
with $SO(3,1)$ isometry.              \par
For case (II), the following two expressions (IIa) and (IIb) are considered: \\
Case (IIa) 
\[
\begin{array}{lll}
A & = (k + \rho)\sin\theta \cosh\omega & {}      
\\
\phi & = (k + \rho) \sinh\omega & {}
\\
\tilde B^1 & = (k + \rho) \cos\theta \cos\varphi \cosh\omega & {}
\\
\tilde B^2 & = (k + \rho) \cos\theta \sin\varphi \cosh\omega. & {}
\end{array}
\]
and \\
Case (IIb) 
\[
\begin{array}{lll}
A &\!\!\! = - (k + \rho) \cos\theta \cos\varphi \cosh\omega &\!\!\! {}
\\
\phi &\!\!\! = (k + \rho) \sinh\omega &\!\!\! {}
\\
\tilde B^1 &\!\!\! = (k + \rho) \sin\theta \cosh\omega &\!\!\ {}
\\
\tilde B^2 &\!\!\! = (k + \rho) \cos\theta \sin\varphi \cosh\omega &\!\!\! {}
\end{array}
\]
%
%
%
%
%
Note that for the case (III) the arguments are the same by exchanging $A$ and $\phi$, 
which we call (IIIa) and (IIIb). 

Substituting these expressions into $ L_{N=2{\rm SUSYQED}}$ $(A, \phi, B^i)$ and 
expanding the action around the vacuum configuration  we obtain the physical particle contents. 
For the cases (IIa) and (IIIa) we obtain 
\begin{eqnarray}
& & 
L_{N=2{\rm SUSYQED}} 
\nonumber \\[.5mm]
& & = 
{1 \over 2} \{ (\partial_a \rho)^2 - 2 (-ef) k^2 \rho^2 \} 
\nonumber \\[.5mm]
& & 
+ {1 \over 2} \{ (\partial_a \theta)^2 + (\partial_a \omega)^2 - 2 (-ef) k^2 (\theta^2 + \omega^2) \} 
\nonumber \\[.5mm]
& & 
+ {1 \over 2} (\partial_a \varphi)^2 
\nonumber \\[.5mm]
& & 
- {1 \over 4} (F_{ab})^2 + (-ef) k^2 v_a^2 
\nonumber \\[.5mm]
& & 
+ {i \over 2} \bar\lambda^i \!\!\not\!\partial \lambda^i 
+ {i \over 2} \bar\chi \!\!\not\!\partial \chi 
+ {i \over 2} \bar\nu \!\!\not\!\partial \nu 
\nonumber \\[.5mm]
& & 
+ \sqrt{-2ef} (\bar\lambda^1 \chi - \bar\lambda^2 \nu) 
+ \cdots, 
\end{eqnarray}
and the consequent mass genaration 
\begin{eqnarray}
& & 
m_\rho^2 = m_\theta^2 = m_\omega^2 = m_{v_a}^2 = 2 (-ef) k^2  
= - {{2 \xi e} \over \kappa}, 
\nonumber \\[.5mm]
& & 
m_{\lambda^{i}} = m_{\chi} =  m_{\nu}= m_{\varphi}= 0. 
\nonumber \\[.5mm]
& & 
%
\end{eqnarray}
(Note that $\varphi$ is NG boson for the spontaneous breaking of $U(1)$ symmetry, i.e. the $U(1)$ phase of $B$,  
and totally gauged away by the Higgs-Kibble mechanism 
for $U(1)$ gauge.) 
The vacuum breaks both SUSY and the local $U(1)$ spontaneously. 
All bosons have the same mass, which is different from the cases (II) and (III) 
and all fermions remain massless, remarkably.
The physical meaning of the off-diagonal mass terms 
$\sqrt{-2ef} (\bar\lambda^1 \chi - \bar\lambda^2 \nu) 
= \sqrt{-2ef} (\bar\chi_{\rm D} \lambda + \bar\lambda \chi_{\rm D})$ for fermions  
is unclear (pathological?), which may induce the mixings of ferm- ions 
and/or the lepton (baryon) number violations.  
%
%

By the similar computations for (IIb) and (IIIb) we obtain 
\begin{eqnarray}
& & 
L_{N=2{\rm SUSYQED}} 
\nonumber \\[.5mm]
& & = 
{1 \over 2} \{ (\partial_a \rho)^2 - 4 f^2 k^2 \rho^2 \} 
\nonumber \\[.5mm]
& & 
+ {1 \over 2} \{ (\partial_a \theta)^2 + (\partial_a \varphi)^2 
- e^2 k^2 (\theta^2 + \varphi^2) \} 
\nonumber \\[.5mm]
& & 
+ {1 \over 2} (\partial_a \omega)^2 
\nonumber \\[.5mm]
& & 
- {1 \over 4} (F_{ab})^2 
\nonumber \\[.5mm]
& & 
+ {1 \over 2} (i \bar\lambda^i \!\!\not\!\partial \lambda^i 
- 2 f k \bar\lambda^i \lambda^i) 
\nonumber \\[.5mm]
& & 
+ {1 \over 2} \{ i (\bar\chi \!\!\not\!\partial \chi + \bar\nu \!\!\not\!\partial \nu) 
- e k (\bar\chi \chi + \bar\nu \nu) \} 
+ \cdots. 
\end{eqnarray}
%
%
and the following mass spectrum which indicates that SUSY is broken spontaneously as expected; 
\begin{eqnarray}
& & 
m_\rho^2 = m_{\lambda^i}^2 = 4 f^2 k^2 = {{4 \xi f} \over \kappa}, 
\nonumber \\[.5mm]
& & 
m_\theta^2 = m_\varphi^2 = m_\chi^2 = m_\nu^2 
= e^2 k^2 = {{\xi e^2} \over {\kappa f}}, 
\nonumber \\[.5mm]
& & 
m_{v_{a}} = m_{\omega} = 0,
\end{eqnarray}
which can produce {mass hierarchy} by the factor 
%
${e \over f}$.  
%
%
The local $U(1)$ gauge symmetry is not broken. 
The massless scalar $\omega$ is a NG boson for the degeneracy of the vacuum in $(A,\tilde B_{2})$-space, 
which is gauged away provided the gauge symmetry between the vector and the scalar multiplet is introduced. 
As for the cosmological significances of $N = 2$ SUSY QED in SGM scenario, 
the vacuum of the cases (IIb) and (IIIb) produces the same interesting predictions 
as already pointed out in $N = 2$ pure SUSY QED in SGM scenario \cite{ST4}, 
which may simply explain the observed mysterious (numerical) relations and give a new insight into the origin of mass 
\begin{center}
$((dark)\ energy\ density\ of\ the\ universe)_{obs} \sim 10^{-12} \sim {(m_\nu)_{obs}}^4 
\sim {\Lambda \over G} \sim {g_{sv}}^2$, 
\end{center}
provided $f \xi \sim O(1)$ and $\lambda^i$ is identified with neutrino. 
($\Lambda$, $G$ and $g_{sv}$ are the cosmological constant of NLSUSY GR (SGM) on {\it empty} new space-time, 
the Newton gravitational constant and the superon-vacuum coupling constant via the supercurrent, 
respectively \cite{KS2,ST4}.)  
While the vacua of the cases (IIa) and (IIIa), though apparently pathological in $d = 2$ so far,  
give new features characteristic of $N = 2$ and may be generic for $N > 2$ and deserve further investigations. 
%
%

The similar investigations in $d = 4$ are urgent and the extension 
of Gell-Mann's idea \cite{GM} to large $N$, especially to $N = 5$, 
is important for {\it superon\ quintet\ hypothesis} in SGM scenario 
with ${N = \underline{10} = \underline{5}+\underline{5^{*}}}$ \cite{KS3} 
and to $N = 4$ is suggestive for the anomaly free nontrivial $d = 4$ field theory.     \par
Also NLSUSY GR in extra space-time dimensions is an interesting problem, 
which can describe all observed particles as elementary {\it a la} Kaluza-Klein. 

Our analysis shows that the vacua of $N$-extended NLSUSY GR action in SGM scenario possess rich stru- ctures 
promissing for the unified description of nature, where $N$-extended LSUSY theory appears 
as the vacuum field configurations of $N$-extended NLSUSY theory on Minkowski tangent space-time.

%
%
%
\newcommand{\NP}[1]{{\it Nucl.\ Phys.\ }{\bf #1}}
\newcommand{\PL}[1]{{\it Phys.\ Lett.\ }{\bf #1}}
\newcommand{\CMP}[1]{{\it Commun.\ Math.\ Phys.\ }{\bf #1}}
\newcommand{\MPL}[1]{{\it Mod.\ Phys.\ Lett.\ }{\bf #1}}
\newcommand{\IJMP}[1]{{\it Int.\ J. Mod.\ Phys.\ }{\bf #1}}
\newcommand{\PR}[1]{{\it Phys.\ Rev.\ }{\bf #1}}
\newcommand{\PRL}[1]{{\it Phys.\ Rev.\ Lett.\ }{\bf #1}}
\newcommand{\PTP}[1]{{\it Prog.\ Theor.\ Phys.\ }{\bf #1}}
\newcommand{\PTPS}[1]{{\it Prog.\ Theor.\ Phys.\ Suppl.\ }{\bf #1}}
\newcommand{\AP}[1]{{\it Ann.\ Phys.\ }{\bf #1}}

\end{document}